\documentclass[prb,amsmath,amssymb,twocolumn,superscriptaddress,showpacs]{revtex4-1}
\usepackage{bm}
\usepackage{amssymb}
\usepackage{colordvi}
\usepackage{graphicx}
\usepackage{color}
\usepackage{hyperref}
\usepackage{ulem}

\newcommand{\beq}{\begin{equation}}
\newcommand{\eeq}{\end{equation}}
\newcommand{\bea}{\begin{eqnarray}}
\newcommand{\eea}{\end{eqnarray}}

\begin{document}
	
	\title{Higher-order topological phases in tunable $C_3$-symmetric photonic crystals}
	
	\author{Hai-Xiao Wang}
	\email{hxwang@gxnu.edu.cn}
	\affiliation{School of Physical Science and Technology, Guangxi Normal University, Guilin 541004, China}

	\author{Li Liang}
	\affiliation{School of Physical Science and Technology, Guangxi Normal University, Guilin 541004, China}

	\author{Bin Jiang}
	\affiliation{School of Physical Science and Technology, \& Collaborative Innovation Center of Suzhou Nano Science and Technology, Soochow University, 1 Shizi Street, Suzhou 215006, China}
	
	\author{Junhui Hu}
	\affiliation{School of Physical Science and Technology, Guangxi Normal University, Guilin 541004, China}
	
	\author{Xiancong Lu}
	\affiliation{Department of Physics, Xiamen University, Xiamen 361005, China}
		
	\author{Jian-Hua Jiang}
	\email{jianhuajiang@suda.edu.cn}
	\affiliation{School of Physical Science and Technology, \& Collaborative Innovation Center of Suzhou Nano Science and Technology, Soochow University, 1 Shizi Street, Suzhou 215006, China}
	
	\date{\today}
	
	\begin{abstract}
	 We demonstrate that multiple higher-order topological transitions can be triggered via the continuous change of the geometry in kagome photonic crystals composed of three dielectric rods. By tuning a single geometry parameter, the photonic corner and edge states emerge or disappear with the higher-order topological transitions. Two distinct higher-order topological insulator phases and a normal insulator phase are revealed. Their topological indices are obtained from symmetry representations. A photonic analog of fractional corner charge is introduced to distinguish the two higher-order topological insulator phases. Our predictions can be readily realized and verified in configurable dielectric photonic crystals.
	\end{abstract}
	
	\maketitle
	
	\section{Introduction}
	Topological phases and phase transitions have been extensively studied in electronic~\cite{Hasan2010,Qi2011}, photonic~\cite{Ozawa2019} and acoustic~\cite{Ma2019,Zhangxj2018} systems in the past decades. Recently, a new class of topological insulators, higher-order topological insulators (HOTIs), which are characterized by higher-order bulk-boundary (e.g., bulk-corner or bulk-hinge) correspondence, were discovered~\cite{Hughes2017Sci,Hughes2017prb,Langbehn2017,Song2017,Schindler2018SCI,Schindler2018Nat,Huber2018,Bahl2018,Imhof2018,AQTI,Hafezi2019,Bernevig2020qudrupole1,Zhenbo2020quadrupole2,Xuyong2019qudrupole3,JJH2020quadrupole,JJH2020polariton,Meng2020quadrupole,Ezawa2018,Zhangbl2018Kagome,Khanikaev2018kagome,LuMH2018dipole,JJH2019Natphys,Hassan2019,Zhangshuang2019,Iwamoto2019dipole,Christensen2019,Xiabz2019elastic,DongJW2019dipole,LuMH2019dipole,JJH2020surfacewave}. Prototype HOTIs include quadrupole and octupole topological insulators~\cite{Hughes2017Sci,Hughes2017prb,AQTI,Huber2018,Bahl2018,Imhof2018,Hafezi2019,JJH2020quadrupole,JJH2020polariton,Zhenbo2020quadrupole2,Meng2020quadrupole,ABK2020octupole,blzhang2020octupole}, three-dimensional HOTIs in electronic systems with topological hinge states~\cite{Langbehn2017,Song2017,Schindler2018SCI,Schindler2018Nat}, and HOTIs with quantized Wannier centers~\cite{Ezawa2018,Noh2018,Zhangbl2018Kagome,Khanikaev2018kagome,LuMH2018dipole,JJH2019Natphys,Hassan2019,Zhangshuang2019,Iwamoto2019dipole,Christensen2019,Xiabz2019elastic,DongJW2019dipole,LuMH2019dipole,blzhang2020thirdorderTI,ABK2020thirdorderTI,JJH2020surfacewave}. For example, quadrupole topological insulators, featured with a fractional bulk quadrupole moment, host gapped edge states with fractional dipole polarization and in-gap corner states accompanying a fractional corner charge. Another example is the $C_3$-symmetric two-dimensional (2D) crystals which exhibit quantized bulk polarization, gapped edge sates and corner states~\cite{Ezawa2018,Zhangbl2018Kagome,Khanikaev2018kagome,Hassan2019,ABKphoton}. In these and other types of HOTIs, the crystalline symmetry plays a key role in the underlying physics. HOTIs set up examples with multidimensional topological physics going beyond the bulk-edge correspondence in conventional topological insulators and semimetals and thus offer novel applications in photonics and phononics.

	Despite the extensive studies on HOTIs, 2D photonic crystals (PhCs) with $C_3$ rotation symmetries are rarely studied~\cite{ABKphoton,Chen2019}. In particular, such PhCs have rich higher-order topological phenomena which have not yet been revealed. Here, we show that by moving the dielectric rods continuously, the $C_3$-symmetric PhCs can switch between triangule and kagome lattice configurations, leading to rich higher-order topological phases and phase transitions. Accompanying with such phase transitions, the corner and edge states emerge or disappear, while the corner charge changes between 0 and $\frac{1}{3}$. The topological indices for various phases are deduced from the symmetry indicators which are closely related to the fractional corner charge~\cite{Hughes2019corcharge}. We also discuss the physical meaning of the fractional corner charge in the photonic context. The richness of the higher-order topological phases and their evolutions provide intriguing photonic phenomena and potential applications in topological photonics which can be readily realized in genuine materials.
	
	\begin{figure}
		\includegraphics[width=3.4in]{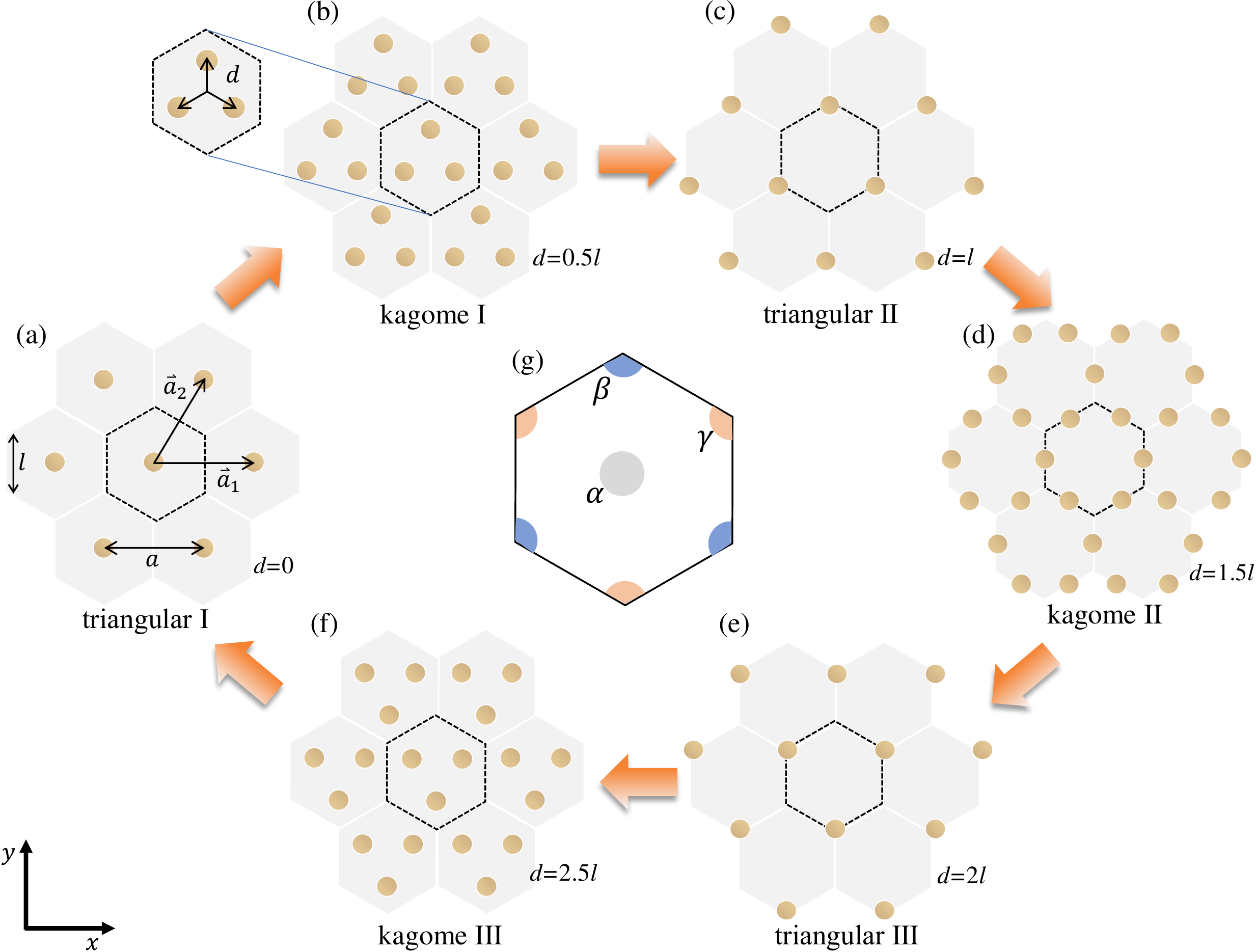}
		\centering
		\caption{(Color online) Geometric transitions in the 2D PhCs with $C_3$ symmetry. The primitive cells are indicated by hexagonal dotted lines with the lattice constant $a$ and the side length $l$. A tunable parameter $d$ with a range of $0$ to $3l$ (the parameter $d$ is modulo $3l$) is employed to illustrate the geometric transitions between triangular, kagome and breathing kagome configurations. By tuning the geometric parameter $d$, the $C_3$ symmetry is preserved, while various configurations can be generated, including (a) triangular I with $d=0$, (b) kagome I with $d=0.5l$, (c) triangular II with $d=l$, (d) kagome II with $d=1.5l$, (e) triangular III with $d=2l$, (f) kagome III with $d=2.5l$. Each primitive cell consists of three dielectric rods (possibly overlaping with each other) with identical radii $r=0.1a$ and permittivity $\epsilon=15$.}
	\end{figure}

	\begin{figure}
		\includegraphics[width=3.4in]{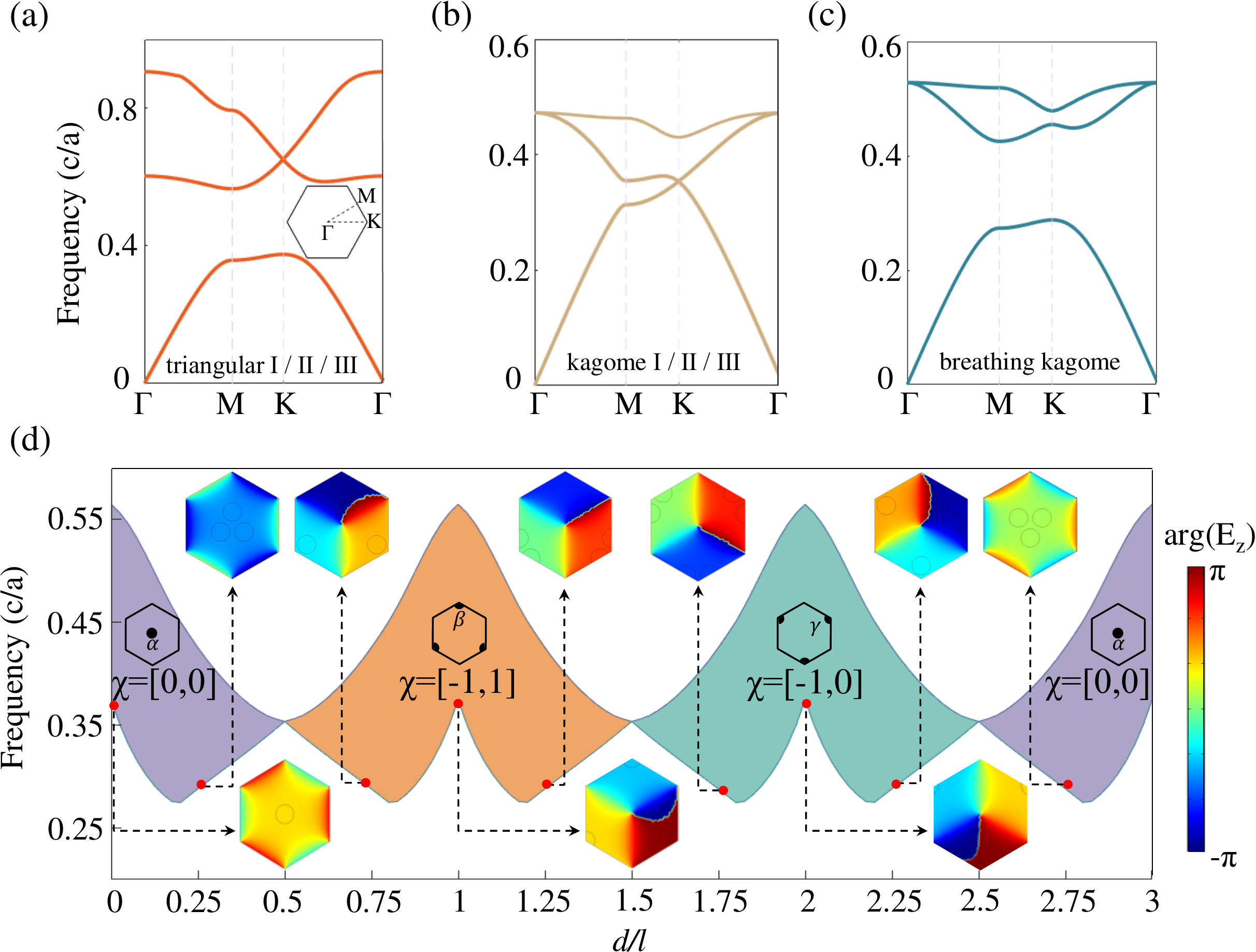}
		\centering
		\caption{(Color online) Photonic band structures of 2D PhCs with $C_3$ symmetry for (a) $d=0, l, 2l$ (i.e., triangular I, II and III lattices), (b) $d=0.5l, 1.5l, 2.5l$ (i.e., kagome I, II and III lattices), (c) $d=0.25l, 1.25l, 2.25l$ (i.e., breathing kagome lattices). (d) The eigen-frequencies of the first and second photonic bands at the $K$ point as functions of $d$. Band gaps of distinct topology are painted with different colors. The topological index $\chi$ is labeled for each region. The Wannier center for each region is depicted as well. Insets illustrate the phase distributions of the eigenstates of the first photonic band at the $K$ point for various $d$'s.}
    \end{figure}

	\section{Higher-order topological phases in tunable $C_3$-symmetric photonic crystals}
	
	We study 2D hexagonal PhCs of $C_3$ rotation symmetry as illustrated in Fig.~1. The lattice vectors are denoted as $\vec{a}_1=(a,0)$ and $\vec{a}_2=(\frac{a}{2},\frac{\sqrt{3}a}{2})$ where $a$ is the lattice constant. The side length of the unit cell is denoted as $l=a/\sqrt{3}$. The simplest configuration is the triangular lattice with a dielectric rod at the center of each unit-cell [Fig.~1(a)] which can be regarded as the special case where three identical dielectric rods overlap with each other. By moving the three dielectric rods along the three symmetry lines, as indicated by the arrows in Fig.~1(b), the PhC undergoes a continuous geometry transformation which includes three triangular lattice configurations (denoted as triangular I, II and III) and three kagome lattice configurations (denoted as kagome I, II and III) [see Figs.~1(b)-1(f)]. The configurations between these six special cases are the breathing kagome lattices. The whole cycle of the continuous deformation encompasses $d$ from 0 to $3l$ (see Fig. 1). The kagome lattices are characterized by $d=(n+\frac{1}{2})l$ with $n=0, 1, 2$, while the triangular lattices are characterized by $d=nl$ with $n=0, 1, 2$.

	Intuitively, as the dielectric rods move, the Wannier center changes. We consider the band gap between the first and the second photonic bands. Therefore, there is only one Wannier center in the unit-cell which can locate at the center ($\alpha$) or the corner of the unit-cell ($\beta$ or $\gamma$) [see Fig.~1(g)]. Unlike the positions of the dielectric rods, the Wannier center is constrained by the crystalline symmetry and thus cannot change continuously. The change of the Wannier center is nonadiabatic which has to be achieved by closing and reopening of the band gap, as shown in details below.
		
	We first provide the photonic band structures for nine prototype cases in Figs.~2(a-c) where we use $c/a$ as the frequency unit ($c$ is the speed of light in vacuum). Throughout this work, we focus on the low-lying photonic bands due to the transverse magnetic (TM) harmonic modes. All the numerical simulation results are carried out by using the finite element numerical solver COMSOL Multiphysics. The photonic bands for the triangular I, II and III configurations are shown in Fig.~2(a) which indicates that the three triangular configurations have identical band structure. This phenomenon is because the three triangular configurations differ only by a partial lattice translation, while the structure of the 2D array of the dielectric rods remain the same. However, the symmetry representations of the photonic bands are distinct for the three triangular configurations. Hence, their topological properties are different. In particular, the location of the Wannier center is distinct for the three triangular configurations, as revealed below.

	Similarly, the photonic band structures for the kagome I, II and III configurations are identical because they can be related to each other by partial lattice translations [Fig.~2(b)]. Such translations change the location of the Wannier center as well as the symmetry representations of the Bloch bands and their topological properties.

	Furthermore, as shown in Fig.~2(d), the photonic band structure is identical, if two configurations differe by an integer times of $l$ in the geometry parameter $d$. Since the lattice periodicity for the translation along the black arrows in Fig.~1(b) is $3l$, there are three different configurations with the same photonic band structure, where the geometry parameter $d$ differs by an integer times of $l$. The translation of $l$ along the black arrows in Fig.~1(b) shifts the unit-cell center to the unit-cell corner without changing the pattern of the 2D array of the dielectric rods. The photonic band structure is insensitive to such global shifts. Therefore, configurations differ by an integer times of $l$ in the geometry parameter $d$ have the same photonic band structure. To further demonstrate such a periodicity of the photonic band structure, we present the photonic bands for three breathing kagome configurations with $d=0.25l$, $1.25l$, and $d=2.25l$ in Fig.~2(c). It is seen that their photonic band structures are identical with each other.

	The evolution of the first two photonic bands at the $K$ point [i.e. ${\bf k}=(\frac{4\pi}{3},0)$] with the geometry parameter $d$ is systematically summarized in Fig.~2(d). There are three topologically distinct photonic band gaps (regions painted by different colors) which are characterized by three different locations of the Wannier center, as indicated in the figure. The band gap between the first two bands experiences closing and reopening during the change of the parameter $d$. We find that the band gap closes at the kagome I, II and III configurations where $d=(n+\frac{1}{2})l$ with $n=0, 1, 2$, separately. These three $d$'s separate the whole region $d\in [0, 3l]$ (bearing in mind that the parameter $d$ is modulo $3l$, since $3l$ corresponds to a lattice periodic translation) into three topologically distinct phases: $d\in (-0.5l, 0.5l)$ where the Wannier center is at $\alpha$, $d\in (0.5l, 1.5l)$ where the Wannier center is at $\beta$, $d\in (1.5l, 2.5l)$ where the Wannier center is at $\gamma$.

	The symmetry representation of the first photonic band at the $K$ point is depicted by the phase profile of the electric field $E_z$ which are shown in Fig.~2(d) for several $d$'s. We find that the $C_3$ symmetry eigenvalue does not change within the same phase. Upon the topological phase transitions, i.e., the band gap closing and reopening, the symmetry eigenvalue changes abruptly. We use the symmetry indicators to characterize the bulk band topology. Following Ref.~\onlinecite{Hughes2019corcharge}, the topological crystalline index can be expressed by the full set of the $C_3$ eigenvalues at the high-symmetry points (HSPs). For a HSP denoted by the symbol $\Pi$, the $C_3$ eigenvalues can only be $\Pi_n=e^{2\pi(n-1)/3}$ with $n=1,2,3$. Here, the HSPs include the $\Gamma$, $K$ and $K^\prime$ points. The full set of $C_3$ eigenvalues at the HSPs are redundant due to the time-reversal symmetry and the conservation of the number of bands below the band gap. The minimum set of indices that describe the band topology are given by~\cite{Hughes2019corcharge},	
	\beq
	[K]_n=\#K_n-\#\Gamma_n, \quad n=1, 2
	\eeq	
	where $\#K_n$ ($\#\Gamma_n$) is the number of bands below the band gap with the $C_3$ symmetry eigenvalue $K_n$ ($\Gamma_n$) at the $K$ ($\Gamma$) point. In this scheme, the $\Gamma$ point is taken as the reference point to get rid of the redundance. For the trivial atomic insulators (i.e., band gap formed by uncoupled atoms), all the HSPs have exactly the same symmetry eigenvalues. Therefore, the trivial atomic insulators have $[\Pi_n]=0$ for all the HSPs. In constrast, any nonzero $[\Pi_n]$ indicates a topological band gap that is adiabatically disconnected from the trivial atomic insulator.

	For the $C_3$-symmetric PhCs, the topological indices can be written in a compact form as
	\beq
	\chi=[[K_1],[K_2]].
	\eeq
	We find that for all $d$'s $\#\Gamma_1=1$ and $\#\Gamma_2=0$. Furthermore, from the $C_3$ eigenvalue at the $K$ point as indicated in Fig.~2(d), we find that the topological index for the three parameter regions are: $\chi=[0,0]$ for $d\in (-0.5l, 0.5l)$, $\chi=[-1,1]$ for $d\in (0.5l, 1.5l)$, and $\chi=[-1,0]$ for $d\in (1.5l, 2.5l)$.

\begin{figure*}
\centering{}\includegraphics[width=0.8\textwidth]{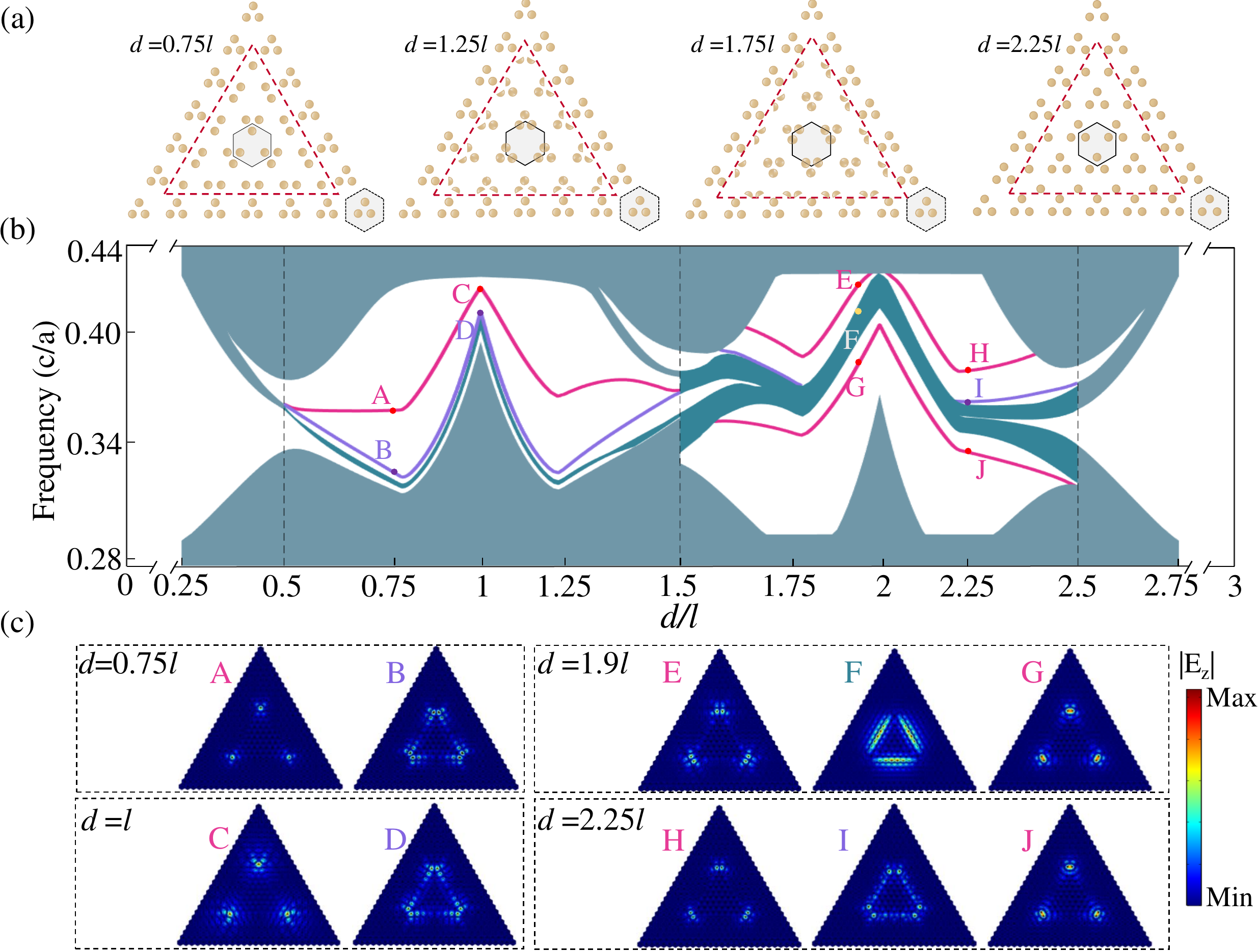}
	\centering
	\caption{(Color online) (a) Schematic illustration of the large triangular supercells with two types of PhCs. The outer PhC has $d=0.25l$, while the inner PhC has variable $d$. Several cases with different $d$'s are shown in (a). (b) Eigen-frequencies of the photons as functions of the geometry parameter $d$. The green-gray regions represent the bulk states, the green regions represent the edge states, while the purple and blue curves represent the type-I and type-II corner states, respectively. (c) Electric field patterns of corner states and edge states with different $d$. Throughout this paper, the electric field patterns of the corner states are given by the superpostion of $|E_z|$ of the three degenerate corner states. In the calculation, the side length of the supercell is $10 a$, while the inside structure has a side length of $4 a$.}
\end{figure*}

\begin{figure*}
\centering{}\includegraphics[width=0.8\textwidth]{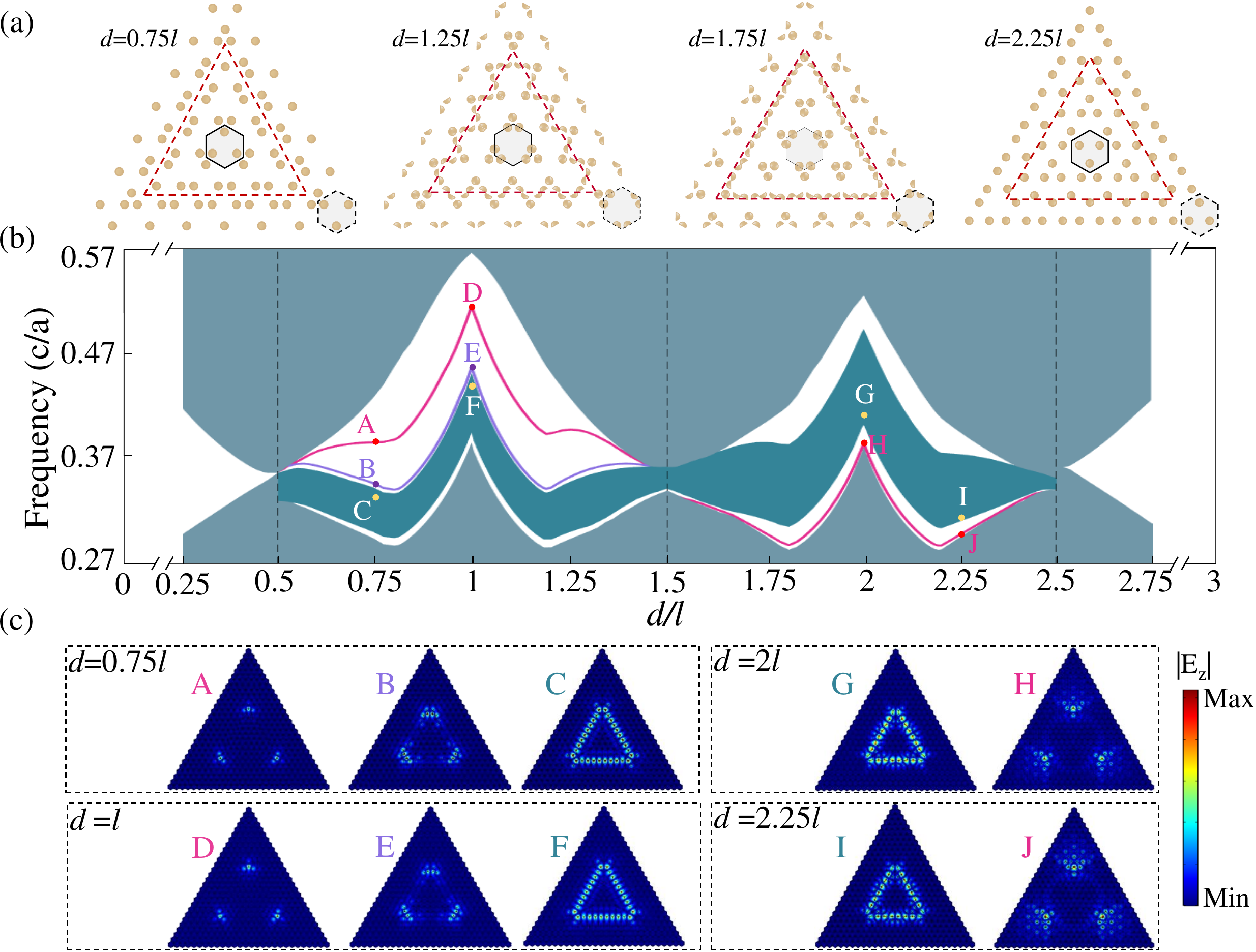}
	\centering
	\caption{(Color online) (a) Schematic illustration of the large triangular supercells with two types of PhCs. The outer PhC has the displacement $d$, while the inner PhC has the displacement $3l-d$. Several cases with different $d$'s are shown in (a). (b) Eigen-frequencies of the photons as functions of the geometry parameter $d$. The green-gray regions represent the bulk states, the green regions represent the edge states, while the purple and blue curves represent the type-I and type-II corner states, respectively. (c) Electric field patterns of corner states and edge states with different $d$. Throughout this paper, the electric field patterns of the corner states are given by the superpostion of $|E_z|$ of the three degenerate corner states. In the calculation, the side length of the supercell is $10 a$, while the inside structure has a side length of $4 a$.}
\end{figure*}

\section{Emergence and evolution of the corner and edge states}
	
	Both the phase with $\chi=[-1,1]$ and that with $\chi=[-1,0]$ are higher-order topological phases that hosting gapped edge states and in-gap corner states. In contrast, the phase with $\chi=[0,0]$ is the trivial phase. To demonstrate the higher-order topology, we construct a large triangular supercell which is schematically shown in Fig.~3(a). In the supercell, the inside is the phase that we study, while the outside is the trivial band gap phase with $d=0.25l$. The side length of the supercell is $10 a$, while the inside structure has a side length of $4 a$. The whole structure is surrounded by the PEC (i.e., perfect electric conductor) boundary condition which is physically a hard-wall boundary for photons.

	We study the evolution of the edge and corner states when the parameter $d$ of the inside PhC structure goes from 0 to $3l$. A number of prototype geometry is shown in Fig.~3(a). The results are presented systematically in Fig.~3(b). Fig.~3(c) gives the electric field $|E_z|$ distributions of the eigenstates. Throughout this paper, the electric field patterns of the corner states are given by the superpostion of $|E_z|$ of the three degenerate corner states. From the figure, it is seen that the edge and corner states emerge only in the region with $0.5l<d<2.5l$, i.e., the two higher-order topological phases. In particular, in the region with $0.5l<d<1.5l$, there are two types of edge states emerging, as revealed previously in Ref.~\onlinecite{ABKphoton}: type-I corner states [denoted by the purple curve in Fig.~3(b), two examples (``A'' and ``C'') are shown in Fig.~3(c)] due to the nearest neighbor couplings and type-II corner states [denoted as the blue curve in Fig.~3(b), two examples (``B'' and ``D'') are shown in Fig.~3(c)] due to long-range couplings. As the common band gap between the inside and outside structures become smaller, at $d=l$, the type-II corner states are much less localized and become edge states alike [see ``D'' in Fig.~3(c)]. In contrast, the type-I corner states remain well-localized and distinguishable from the edge states [denoted as green in Fig.~3(b)] and bulk states [denoted as green-gray in Fig.~3(b)]. This indicates that the type-I corner states are due to the bulk topology, while the type-II corner states may originate from long range couplings between the adjacent edge states.

	The region with $1.5l<d<2.5l$ has not yet been studied in the literature. We find that in this region the type-II corner states are hardly seen. Meanwhile, there are two sets of type-I corner states. Each set has three degenerate corner states. One set has frequency higher than the edge states, while the other set has frequency lower than the edge states. In both sets, the wavefunctions of the corner states are well-localized around the corners, distinghuisable from the edge states (the green band) and the bulk states (the green-gray bands). Examples of the corner and edge wavefunctions at $d=1.9l$ and $d=2.25l$ are shown in Fig.~3(c). In some cases, for instance, $d=2.25l$, type-II corner states can be found. However, the wavefunctions are not well-localized at the corners.
    
	We now explore the corner and edge states in another type of supercell. We design the supercell in such a way that the inner structure is a PhC with the parameter $d$ while the outer structure is the PhC with the parameter $3l-d$. That is, we consider the edge and corner boundaries between complimentary PhC structures. Such a supercell architecture will induce intriguing edge and corner boundaries [e.g., various zigzag edge boundaries as depicted schematically in Fig.~4(a)]. The evolution of the bulk, edge and corner states are systematically summarized in Fig.~4(b) and 4(c).

	In this type of supercells, the edge and corner states emerge only in the two topological regions, $0.5l<d<1.5l$ and $1.5l<d<2.5l$, as shown in Fig.~4(b). In the region with $0.5l<d<1.5l$, the inner PhC has the topological index $\chi=[-1,1]$ while the outer PhC has the topological index $\chi=[-1,0]$. In the other region with $1.5l<d<2.5l$, the topological indices of the inner and outer PhCs switch, i.e., the inner PhC has $\chi=[-1,0]$ and the outer PhC has $\chi=[-1,1]$. The emergence of the edge and corner states, which has not yet been discovered in the literature, reveals that higher-order topological phenomena can appear at the boundaries between two topologically distinct higher-order phases. This is consistent with the topological band theory~\cite{Hughes2019corcharge,JJH2020invsym} and the Wannier center picture.

	From Fig.~4(b), the bulk band gap closing is clearly seen at the phase transition points, $d=0.5l$, $1.5l$ and $2.5l$. Type-II corner states can be found only in the higher-order phase with $\chi=[-1,1]$ (i.e., the phase studied in Ref.~\onlinecite{ABKphoton}; here $0.5l<d<1.5l$), but not in the higher-order phase with $\chi=[-1,0]$ (i.e., $1.5l<d<2.5l$). For all cases in the region $0.5l<d<2.5l$, the edge states are clearly visible [see Fig.~4(c)]. The bandwidth of the edge states is considerably larger in this type of supercells compared with the supercell studied in Fig.~3. As a consequence, the corner states are less localized, particularly in the region with $1.5l<d<2.5l$ where the corner states live in the small band gap between the edge and bulk states.

\begin{figure}
	\includegraphics[width=3.4in]{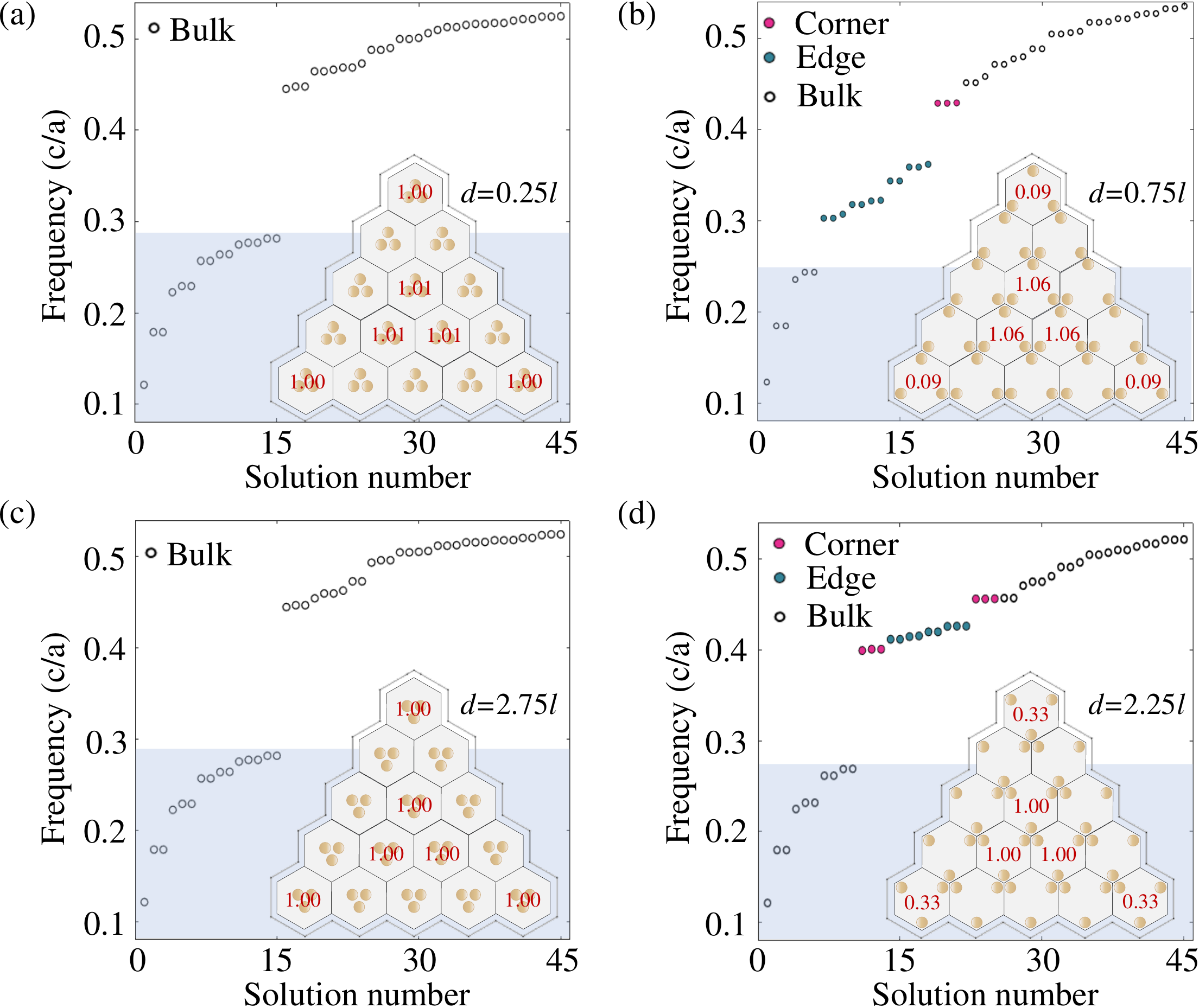}
	\centering
	\caption{Fractional ``charges'' in the triangular supercell with perfect electric conductor boundary conditions. Only the charges of the corner unit-cells and the bulk unit-cells are shown in the figure, as indicated by the blue areas. Four cases are considered (a) $d=0.25l$, (b) $d=0.75l$, (c) $d=2.75l$ and (d) $d=2.25l$. }
\end{figure}

\section{fractional corner charge}

	We now show that the higher-order band topology can also be manifested in the fractional corner charge. Even though we are considering photonic bands and photonic states in this work, it is possible to define an analog of ``charge'' through the local density of states (LDOS), $\rho_e({\bf r}, E)$. In electronic systems, the charge contributed by the filling of the valence bands in the $j$-th unit-cell is given by
	\beq
	Q_{j,e} = e \int^{E_{\rm gap}} d E \int_{j} d{\bf r} \rho_e({\bf r}, E) .
	\eeq
	Here, $e$ is the charge of an electron, $E_{\rm gap}$ is an energy in the topological band gap which is below the eigen-energies of the edge and corner states. In the above equation, the integration over the position ${\bf r}$ is defined within the $j$-th unit-cell. The filling of all the valence bands below the topological band gap contributes a fractional charge in the corner region. It was predicted~\cite{Hughes2019corcharge} that the fractional corner charge $eQ_c$ is completely determined by the topological indices of the bulk bands as follows,
	\beq
	Q_c = - \frac{1}{3}([K_1]+[K_2]) \mod 1 .\label{qc-eq}
	\eeq
	The actual size of the corner region depends on the specific model. However, one can often choose one or a few unit-cells around the corner boundary to converge the fractional corner charge.

	We then check the theoretical prediction in photonic system by calculating the following quantity which is the analog of the ``charge'' in the $j$-th unit-cell in the photonic system,
	\beq
	Q_j = \int_0^{f_{\rm gap}} df \int_{j} d{\bf r} \rho_p({\bf r}, f) .
	\eeq
	Here, we omitted the elementary charge $e$ which does not have a physical meaning in photonics. The integration over frequency is from 0 to a frequency in the band gap $f_{\rm gap}$ which is below the eigen-frequencies of the edge and corner states. The photonic LDOS is calculated through the following spectral decomposition of all the photonic eigenstates of the valence bulk bands
	\beq
	\rho_p({\bf r}, f) = \sum_{n} \frac{\Gamma}{\pi}\frac{\epsilon({\bf r})|E_z^{(n)}({\bf r})|^2}{(f-f_n)^2+\Gamma^2} .
	\eeq
	Here, $n$ labels the photonic eigenstates of the valence bulk bands, and $\Gamma\to 0$ is a sufficiently small number that converges the calculation. $E_z^{(n)}({\bf r})$ is the scaled electric field distribution of the $n$-th photonic eigenstate which satisfies the following normalization condition
	\beq
	1= \int d{\bf r} \epsilon({\bf r})|E_z^{(n)}({\bf r})|^2
	\eeq
	where the integral of ${\bf r}$ is over the whole photonic system and $\epsilon({\bf r})$ is the position-dependent relative permittivity.

	The photonic ``charge'' defined above does have a physical meaning. It represents the number of the photonic modes contributed from the $j$-th unit-cell from the valence bulk bands. We calculate the photonic ``charge'' for each unit-cell and present the results in Fig.~5 for various configurations.

	For all the four cases considered in Fig.~5, the calculated charge for the bulk unit-cells are close to 1. This is consistent with the fact that there is only one band below the band gap, i.e., each unit-cell contributes a single charge (mode) to the bulk band. Figs.~5(a) and 5(c) show that for both $d=0.25l$ and $d=2.75l$ (i.e., $\chi=[0,0]$), the fractional corner charge is zero, since the corner unit-cell has a charge very close to 1. Fig.~5(b) shows that for $d=0.75l$ (i.e., $\chi=[-1,1]$), the fractional corner charge is $Q_c=0$ which is indicated by that the corner unit-cell has a charge very close to $0$. In this case, despite that the band gap carries higher-order topology and the resultant corner states, the corner charge vanishes, which is consistent with the theoretical prediction given in Eq.~(\ref{qc-eq}). Fig.~5(d) shows that for $d=2.25l$ (i.e., $\chi=[-1,0]$), the fractional corner charge is $Q_c=1/3$ which is manifested by that the corner unit-cell has a charge very close to $1/3$. These photonic charges can be measured through the classical or quantum versions of the Purcell effect, as indicated by Ref.~\onlinecite{disclination}.

\section{Conclusion}
	In conclusion, we demonstrate that rich higher-order topological phases and multiple phase transitions can be obtained in $C_3$ symmetric PhCs by tuning a single geometry parameter $d$. These higher-order topological phases yield intriguing multidimensional topological phenomena where the corner and edge states can be tuned in versatile ways. Our study shows that continuously configurable dielectric PhCs~\cite{Zhu2018} can be useful in generating topological photonic circuits with tunable edge and corner states. The emergent fractional photonic charge indicates that photonic systems can be powerful in revealing the fundamental properties of topological bands.

	\begin{acknowledgements}
	H.-X. Wang and L. Liang contributed equally to this work. This work is supported by the National Natural Science Foundation of China under Grant No. 11904060, 12074281. J.-H. Jiang is supported by the Jiangsu specially-appointed professor funding, and a project funded by the Priority Academic Program Development of Jiangsu Higher Education Institutions (PAPD).
	
	\end{acknowledgements}

\end{document}